# Oxidation of graphene on metals

Elena Starodub[*†], Norman C. Bartelt, and Kevin F. McCarty[†]

Sandia National Laboratories, Livermore, CA 94550

We use low-energy electron microscopy to investigate how graphene is removed from Ru(0001) and Ir(111) by reaction with oxygen. We find two mechanisms on Ru(0001). At short times, oxygen reacts with carbon monomers on the surrounding Ru surface, decreasing their concentration below the equilibrium value. This undersaturation causes a flux of carbon from graphene to the monomer gas. In this initial mechanism, graphene is etched at a rate that is given precisely by the same non-linear dependence on carbon monomer concentration that governs growth. Thus, during both growth and etching, carbon attaches and detaches to graphene as clusters of several carbon atoms. At later times, etching accelerates. We present evidence that this process involves intercalated oxygen, which destabilizes graphene. On Ir, this mechanism creates observable holes. It also occurs mostly quickly near wrinkles in the graphene islands, depends on the orientation of the graphene with respect to the Ir substrate, and, in contrast to the first mechanism, can increase the density of carbon monomers. We also observe that both layers of bilayer graphene islands on Ir etch together, not sequentially.

## 1. Introduction

How graphitic carbon reacts with oxygen is important to many diverse fields such as combustion [1] and the regeneration of catalysts poisoned with carbon [2]. Owing to the promising electronic properties of graphene [3], how oxidation creates defects in graphene has recently received considerable attention [4, 5, 6, 7]. In this paper, we address the question of how oxygen etches graphene on Ru(0001) and Ir(111) as CO forms and desorbs.

---

[*] Formerly Elena Loginova.
[†] To whom correspondence should be addressed. E-mail: mccarty@sandia.gov (K.F.M.); estarod@sandia.gov (E.S.)

Our previous work has established that graphene islands on Ru(0001) and Ir(111) are in equilibrium with a relatively dense gas of C monomers on the surrounding metal surface at the temperatures where high-quality films grow [8-12]. The equilibrium occurs by a constant exchange of carbon atoms between the graphene and the surrounding adatom gas. The existence of this equilibrium suggests two scenarios of how graphene islands might be etched by $O_2$ exposure. In one limit, illustrated in Figure 1a, oxygen quickly dissociates on the Ru, reacts with the adatoms to form CO, which desorbs. This process depletes the adatom concentration. In this "Langmuir-Hinshelwood" mechanism [13], the graphene is etched because of a net evaporation rate into the adatom gas to restore its equilibrium concentration. In the other limit, (dissociated) oxygen directly attacks the island, as sketched in Figure 1b, bypassing the adatom gas. Recent photoelectron emission microscopy observations of graphene etching during oxidation on Ru(0001) did not distinguish these mechanisms [14].

Here we present in-situ observations of graphene oxidation revealing that both these mechanisms can occur, depending on the oxygen dose. We present evidence suggesting that the direct-attack mechanism (Figure 1b) involves oxygen intercalation under graphene. In addition, we show that the in-plane orientation of the graphene sheet with respect to the substrate greatly affects etching kinetics. We show that both layers of bilayer islands are etched together, as opposed to layer-by-layer. Finally, wrinkles where graphene has locally debonded from the substrate are found to etch significantly faster than adjacent regions.

2. **Results and discussion**

**2.1. Graphene etching on Ru(0001).** To address which of the two oxidation mechanisms shown in Figure 1 occurs on Ru(0001), we use low-energy electron microscopy (LEEM) to monitor the decrease in adatom density during oxygen exposure and then correlate this density with the graphene etching rate. In similar experiments during graphene growth, we established that the velocity $v$ at which a graphene step edge advances depends on the adatom concentration $c$ through [8]:

$$v = B\left[\left(\frac{c}{c^{eq}}\right)^n - 1\right], \quad (1)$$

where $B$ is a temperature-dependent constant, $n$ is the number of carbon adatoms that attach in the rate-limited step of growth, and $c^{eq}$ is the adatom concentration in equilibrium with graphene. We found n = 4.8 ± 0.5 for growth on Ru(0001) [8, 10] and n ≈ 5 for the most common orientation of graphene on Ir(111) [15], implying that carbon is exchanged between the graphene and the adatom gas in units of ~5 carbon atoms. The nature of this exchange should not change dramatically when $c$ is decreased slightly below $c^{eq}$, causing etching ($v < 0$). Thus, if eq 1 is valid for $c < c^{eq}$ during oxygen exposure, oxidation occurs through the adatom gas.

To test this hypothesis, we used LEEM to monitor the size of monolayer graphene islands on Ru(0001) during oxidation at temperatures between 980 to 1070 K, and monolayers and bilayers on Ir(111) [8]. Figure 2 summarizes a typical experiment. The left panel shows LEEM images of the Ru(0001) surface (grey) with a graphene island (dark) decreasing in size during exposure to 5×10$^{-9}$ Torr $O_2$. In the images, the intensity (contrast) of the island remains constant and spatially uniform. The island completely disappears after 274 s. For temperatures from 980 - 1070 K and for $O_2$ pressures below $5\times10^{-7}$ Torr, we observed no holes forming within individual graphene islands on Ru(0001). Under these conditions, we conclude that etching occurs primarily at or near island edges.

From the LEEM images we determine island area and perimeter, as shown in the lower right panel of Figure 2. From the intensity $I(t)$ of the electron beam specularly reflected from a graphene-free region [16], we also simultaneously determine the concentration of mobile carbon adatoms $c$ in monolayers (ML) using [8]:

$$c(ML) = c^{eq} + 0.223 \times \frac{I^{eq} - I(t)}{I^{eq}}. \qquad (2)$$

Here, $I^{eq}$ is the reflected intensity from a graphene-free region with an equilibrium adatom concentration $c^{eq}$, as measured before oxygen exposure (the red curve in the top right panel of Figure 2). The calibration factor of 0.223 was determined in Ref. [8]. Since oxygen is removed from the graphene-covered Ru(0001) by the formation and desorption of CO at temperatures above 950 K and oxygen pressures below $1\times10^{-8}$ Torr [17], it is unlikely that oxygen changes the surface reflectivity during oxidation at our etching temperatures. (Oxygen adsorption does change the reflected intensity after all carbon is removed, however.)

Before oxygen is introduced, the graphene in Figure 2 is in equilibrium with an adatom concentration $c^{eq}$ = 0.017 ML, and island areas do not change with time [8]. In $5\times10^{-9}$ Torr of oxygen, the adatom concentration decreases as carbon adatoms react with oxygen to produce CO, which then desorbs [17, 18]. From an island's area $A$ and perimeter $P$, we calculate the speed of the island edge $\upsilon = dA/Pdt$, as shown in the insert of the bottom right panel of Figure 2. The edge velocity becomes negative upon oxygen introduction. At ~175 s, the C adatom concentration drops to a small fraction of the initial value and the step velocity accelerates greatly. The velocity continues to increase with time, reaching over 80 nm/s at 250 s.

Knowing the separate time dependencies of the edge velocities and the carbon adatom concentrations (Figure 2), we can determine $v$ as of function of $c$ and check if eq 1 is satisfied.‡ The open circles in Figure 3 plot the edge velocities vs. C adatom concentrations during oxidation at three temperatures. We set the adatom concentrations prior to oxygen exposure to be the $c^{eq}$ values previously measured [8]. As oxygen exposure removes carbon adatoms, their concentration falls below $c^{eq}$ and the edge velocity becomes negative. Initially the step velocity increases only slowly as the adatom concentration decreases. The filled circles in Figure 3 plot the same velocity/concentration relationship during *growth* from C deposition at the same temperatures, as reported in reference 7. The black lines in Figure 3 show the previously reported fits of the growth data to eq 1 extrapolated to etching conditions (i.e., $c < c^{eq}$). Remarkably, the slow etching rates (velocities below 1 nm/s) close to $c^{eq}$ are accurately described by the growth-data fits. In particular, the fits to the growth data predict that the etching velocity in the asymptotic limit as the carbon monomer concentration becomes small should increase from -0.30 ± 0.08 nm/s to -2.62 ± 0.06 nm/s as temperature increases from 980 K to 1070 K. Before the etching accelerates at late times as discussed below, the small-concentration limits of the experimental data are within a few tenths of a nm/s of these predicted changes, as shown in Figure 3. The verification of this prediction increases our confidence in the validity of eq 1 and shows that the rate-limiting step of both graphene etching and growth involve the same clusters of ~5 carbon atoms.

The etching kinetics in Figure 3 only follows the same functional relationship (eq 1) as the growth kinetics during the initial stage of oxidation. At later times, the oxidation rate accelerates as the C adatom concentration decreases, deviating dramatically from the fits of eq 1.

---

‡ That oxygen pressure varies with time in our experiments does not complicate this check because none of the parameters of eq 1 depend on oxygen pressure.

Clearly etching is now occurring by a different mechanism. Some insight into this second, fast mechanism comes from examining how its onset depends on oxygen pressure, as shown in Figure 4a for five different $O_2$ pressures at 1020 K. The onset does not occur at a fixed adatom concentration – indeed as pressure decreases, fast etching occurs even though the depletion of the adatom gas is small. However, the oxygen exposure needed to initiate the second (fast) mechanism is approximately the same (~0.7 L) for all the oxygen pressures, as shown in Figure 4b. This fixed exposure suggests that the onset of fast oxidation occurs after filling an oxygen reservoir to a critical value. Assuming an oxygen sticking coefficient of one, the total critical coverage is around ~0.02 ML for a surface covered with ~0.06 ML graphene. We hypothesize that this reservoir is O intercalated underneath graphene, which allows the oxygen to directly react with the graphene as in Figure 1a. We next provide evidence consistent with this proposal of direct attack by intercalated oxygen.

**2.2. The attack of graphene by intercalated oxygen.** At lower temperatures oxygen can be adsorbed on a Ru surface without reacting with the graphene to form CO. Figure 5 shows LEEM images of a graphene island under these conditions (550 K and $6\times10^{-8}$ Torr oxygen). Initially, electron reflectivity from the Ru surface decreases. After ~130 seconds, the reflectivity stops changing. (When this saturated surface is cooled to room temperature, half-order low-energy electron diffraction (LEED) spots characteristic of the 2×1 phase of O on Ru(0001) appear [19].) Further oxygen exposure results in a band of different contrast at the island perimeter. This band expands towards the center of the island. Before it reaches the center, however, a second change in contrast appears at the island perimeter when the oxygen dose exceeds ~40 L. This region then expands, following the first region towards the island center. After ~3000 seconds, the island is completely converted to a phase with the second type of contrast.

LEED from the center of the island contains two, distinct 6-fold patterns of the same orientation but different spacing. This result shows that the graphene and Ru lattices diffract independently and that their unit cell vectors are parallel. In contrast to this evidence of weak film/substrate interaction, the initial (non-intercalated) graphene on Ru(0001) has pronounced superstructure diffraction spots due to strong substrate/film interaction [20]. Evidently this structural change results from dissociated O intercalating under the graphene weakening its interactions with the substrate. Our LEEM and LEED observations are consistent with the results of Zhang et al. [21], who found that low-temperature oxygen exposure removed the corrugation in scanning tunneling microscopy that is characteristic of the strong graphene/Ru(0001) interaction. The occurrence of two well-defined fronts during the intercalation process shows that it occurs in stages (Figure 5), perhaps through progressively higher concentrations of underlayer oxygen.

After stopping exposure to oxygen, but maintaining the intercalation temperature, the diffraction from the island perimeter has the non-intercalated LEED pattern (i.e., contains moiré-type superstructure spots). Perhaps removing the intercalated O around the island rim allowed the graphene there to become more tightly bound to the Ru, sealing the island interior to further oxygen loss. To determine how the intercalated oxygen reacts with the graphene, we slowly heated the sample in vacuum, as shown in Figure 6. The image focus used in Figure 6 reveals that the starting intercalated graphene (0-s image) is covered with a labyrinth of dark, linear defect features. At 720 K, the center of the graphene dramatically disintegrates within a few seconds. Only an outer ring of graphene is left, where we suggested the intercalated oxygen had previously escaped. Accompanying this "explosion" is a complex sequence of changes in electron reflectivity away from the island, presumably caused by carbon and oxygen being released by the disintegrating island and reacting to form desorbing CO. The sudden decomposition illustrates that oxygen-intercalated graphene is highly unstable.

This experiment clearly shows that intercalated oxygen can directly attack graphene at an extreme rate at temperatures as low as 720 K (Figure 6). A consistent explanation of the higher-temperature oxidation in Figures 2-4 is that oxygen intercalates only a small distance under the graphene before oxidation occurs. The resulting narrow band of intercalated graphene would not be discernable in the LEEM images, explaining why the graphene island in Figure 2 does not change contrast, unlike in Figure 5. Taken together, these observations are consistent with our hypothesis that a form of direct attack by intercalated oxygen is responsible for the second (fast) oxidation mechanism on Ru. That this direct attack only occurs after a critical exposure suggests that a fixed amount of intercalated O is required to initiate this etching mechanism. Since all of the etching we observe before this critical exposure is reached is caused by oxygen reaction with the monomer gas, it appears that simple direct attack of carbon at the edges of non-intercalated graphene does not occur.

To summarize, oxygen etches graphene on Ru(0001) through two mechanisms. The initial oxygen exposure removes C adatoms, which are replenished by C from the graphene. This Langmuir-Hinshelwood-type mechanism of etching (Figure 1a) follows the same cluster addition/subtraction kinetics as graphene growth (eq 1 and Figure 3). The etching mechanism changes when a certain amount of oxygen has been dosed onto the surface (Figure 4b). We also find that intercalated oxygen oxidizes graphene at an extremely high rate (Figure 6). These two observations suggest that the second, fast mechanism occurs by the direct attack of intercalated oxygen on the graphene (Figure 1b) [§]. We next contrast these observations to oxidation on Ir(111) where the fast, direct-attack mechanism dominates. The detailed behavior depends upon specific characteristics of the graphene, including its orientation on the Ir substrate and the presence of delaminated wrinkles.

**2.3. Etching of rotational variants on Ir(111).** The etching of graphene on Ir(111) is complicated by the existence of at least four rotational variants that differ in their in-plane orientation with respect to the substrate [11]. The rotational variant we label R0 has the same film/substrate orientation as always observed on Ru(0001) [8, 10, 22]. The other variants, R30, R18.5 and R14, are rotated by 30°, 18.5° and 14° [11]. Figure 7 describes the oxidation on Ir(111) using the same methodology as above. The LEEM images in the left panel show two abutting islands of R0 (dark) and R30 (bright) orientation, respectively. Similar to graphene on Ru(0001), the contrast of the graphene islands remains invariant during etching at this low oxygen pressure of $1.6 \times 10^{-9}$ Torr. No obvious holes form in the graphene. As on Ru, we conclude that graphene on Ir(111) at low oxygen pressure is being etched near the graphene edges. However, the LEEM images in Figure 7, taken at equal time intervals, reveal markedly different etching kinetics for the two rotational variants – the R30 island is etched much faster than the R0 island, consistent with the results of Van Gastel et al.[23]. The lower right panel of Figure 7 plots the step velocity of the R0 and R30 islands. The step edge of the R30 island retracts at a relatively constant velocity as soon as the $O_2$ exposure starts. In contrast, the R0 island is hardly etched until the R30 island is gone. After the R30 island vanishes, the step velocity of the R0 island becomes progressively faster, in contrast to the nearly constant step velocity of the R30 domain.

---

[§] The activation energy in the first mechanism should be the same as in growth, i.e., 2 eV [8]. At generally higher pressures and lower temperature, Cui et al. [14] measure a much lower activation energy of ~0.3 eV. Thus, the etching of Cui et al. likely occurs by the direct attack mechanism.

We also measured the C adatom concentration on Ir(111) during etching from the change in electron reflectivity [15]. The R0 and R30 graphene islands are at equilibrium with the same density of C monomers before oxygen is introduced. (This is evidence that the two rotational domains have similar energies [9].) The upper right panel in Figure 7 shows the deviation from this initial equilibrium concentration during oxygen exposure. Surprisingly, the monomer concentration *increases* when oxygen is introduced.** The adatom concentration reaches 0.010 ML above the equilibrium value. Only when the R30 island vanishes, at about 2200 s, does the adatom concentration start to decrease. The remaining R0 island then begins to be oxidized at progressively faster rates, as seen in the lower right panel of Figure 7. During the oxygen exposure, both the R0 and R30 islands are being etched even when the C adatom sea is supersaturated. Clearly neither rotational variant is being oxidized by adatom detachment driven by adatom under-supersaturations, the Langmuir-Hinshelwood-type mechanism found at early times on Ru (Figure 1a). Instead, as we discuss in Sections 2.4 - 2.6, there is significant evidence that intercalated oxygen is responsible, as for Ru.

Why should the carbon adatom concentration increase during oxidation? One possibility comes from considering point defects created by reaction with oxygen in the interior of graphene islands. The lowest-energy defects in graphene cost several eV [24]. In contrast, the energy of forming a monomer from a graphene edge is less than 0.5 eV [8, 9]. So only a low density of point defects would be needed to make the intact graphene higher in energy than a monomer gas with the same number of carbon atoms. Thus, if the defect creation rate is much greater than the healing rate from monomer reattachment, the island will decompose to monomers. From our growth experiments, the supersaturation we observe during etching are not great enough to nucleate new graphene islands at an appreciable rate.

---

** We note that oxygen exposure without surface C had no measureable effect on the electron reflectivity from Ir(111).

Van Gastel et al. have used the faster etching of the R30 variant to selectively remove it from graphene films on Ir(111) [23]. This fast etching kinetics of the R30 variant mirrors its fast growth kinetics. That is, the R30 graphene grows markedly faster than R0 graphene at the same supersaturation [9].

**2.4. Hole formation at high oxygen pressures on Ir(111).** While no holes formed in graphene on Ir(111) for the oxygen pressure of Figure 7, we did observe them at higher pressures. Figure 8 shows an example during oxidation at twice the pressure, $4\times10^{-9}$ Torr. The 0-s image shows two abutting islands. The upper graphene (dark) is R0 type and the lower region (bright) is R30 type. Upon $O_2$ exposure, bright patches occur in the 90-s image. We interpret these patches to be holes that expose the substrate. Holes then begin to form in the R0 domain, starting near the domain perimeter, as seen in the 125-s image. The hole formation starting near the island edges is consistent with oxidation being preceded by intercalation of oxygen under the graphene. We do not observe hole formation on Ru(0001) for pressures over 100 times that of Figure 7. (See section 2.1.)

The hole formation suggests that oxygen can penetrate more easily under graphene in the case of the Ir(111) substrate, leading to the mechanism of direct attack by intercalated oxygen. This intriguing finding is consistent with the theoretical calculations by N'Diaye et al. [25] and Martoccia et al. [22], which showed that graphene is more weakly bound to Ir(111) than to Ru(0001), thus making intercalation easier. One explanation of the critical oxygen dose required for direct oxidation of graphene on Ru (Figure 4b) is that a critical amount of intercalated oxygen is required to remove the strong binding with the substrate [21], allowing even more intercalated O and the initiation of the faster etching by direct attack. On Ir, direct attack occurs almost immediately, bypassing the Langmuir-Hinshelwood-type mechanism of Figure 1a.

**2.5. Etching of bilayer graphene on Ir(111).** We next consider the oxidation of graphene bilayers on Ir(111). Figure 9 shows the etching of R30 graphene in $6\times10^{-7}$ Torr of $O_2$. Most of the film is one layer thick, which images bright grey. The dark island in the 0-s image, however, is two-layer graphene, as confirmed by measuring oscillations in electron reflectivity vs. electron energy and selected-area low-energy electron diffraction. The graphene is etched from top to bottom in the images. As in Figure 8, holes (dark) occur in the island near its horizontal boundary with the bare substrate. The holes occur up to about 2 μm away from the sheet edge, showing that oxidation is occurring by direct attack.

After 20 s of oxygen exposure, the etching front reaches the two-layer island in Figure 9. The single-layer graphene is etched from around the multilayer island, as seen in the 34- and 48-s images. Since the contrast of the etched bilayer island remains unchanged, we conclude that both graphene layers are etched simultaneously. The schematic of the etching process is shown on the right panel. This observation would be consistent with the bottom layer being removed by intercalated O, and once the bottom layer is removed, the upper layer is subject to immediate attack by the same process.

Inspection of videos of the etching process shows that the edge of the single-layer graphene becomes pinned (bowed) when passing though defects such as the bilayer island in Figure 9. But once freed from a pinning site, a straight etching front is quickly reestablished. This morphological stability of the graphene edge suggests that the etching process is diffusion-limited under the conditions of Figure 9 [26]. In the diffusion-limited regime of etching, a section of graphene edge that is bowed towards the bare substrate is closer to the source of oxygen on the Ir terrace and etches faster until it catches up with the rest of the retreating edge.[††]

**2.6. Etching of graphene ridges on Ir(111).** Finally we consider how oxygen etches wrinkles in graphene on Ir(111). These wrinkles form during cooling from the growth temperature due to the large thermal contraction of the substrate relative to graphene [11, 27]. (We do not observe wrinkle formation on Ru(0001), presumably because of graphene's stronger interaction with this substrate [20].) Graphene was grown on Ir(111) at 1150 K and then cooled down to 300 K to form wrinkles. Figure 10 shows two abutted, wrinkled islands. The top island is an R0 variant while the bottom one is an R30 variant [11]. In the upper, pre-exposure image, the most visible wrinkles lie roughly along the vertical direction. The islands were then exposed to $2 \times 10^{-9}$ Torr of oxygen at 670 K. Bright spots and lines are present in the islands after 48 min of exposure (middle image). We interpret these bright regions as holes in the graphene islands. Typically, these holes occur at the wrinkles, starting from the island edges. Notice, however, wrinkles are sometimes oxidized even if they have no direct connection to the graphene edge. The wrinkles are completely etched after 78 min of oxygen exposure, forming gaps in the islands. The island edges are only slightly oxidized over the same time period as the island size changes little. These observations suggest that the wrinkles are also etched by the second, fast mechanism represented in Figure 1b. Since they are debonded from the substrate and oxygen should easily intercalate under them, attack by intercalated oxygen again seems likely.

**3. Conclusions**

---

[††] Carbon atoms diffusing along the edge of the graphene sheet could also smooth edge roughness. However, we see no evidence of significant edge diffusion since graphene islands maintain highly non-equilibrated shapes during growth and annealing at even higher temperatures.

The mechanisms by which oxygen etches a single layer of graphite on a metal substrate have been elucidated. On Ru, oxidation proceeds by two sequential mechanisms. The first involves oxygen interaction with the carbon adatom sea on the metal surface, a Langmuir-Hinshelwood process. This first, slow mode of graphene etching confirms the remarkable mechanism of graphene growth limited by 5-atom cluster attachment. The second mechanism likely involves oxygen penetration underneath the graphene. We find that oxygen intercalated under graphene on Ru(0001) is remarkably unstable. The in-plane orientation of the graphene has a large affect on the oxidation rate, consistent with our previous observations that orientation greatly influences the growth rate [9]. We observe that both layers of bilayer islands etch together, not sequentially. Finally, linear ridges where the graphene had locally debonded from the substrate during cooling were found to oxidize substantially faster than flat graphene.

**Acknowledgment.** The authors acknowledge with pleasure a valuable discussion with Alex Chernov (LLNL) about mechanisms of crystal growth and etching. This work was supported by the Office of Basic Energy Sciences, Division of Materials Sciences and Engineering of the US DOE under Contract No. DE-AC04-94AL85000.

**Referenced and Notes**

(1.   Whitesides, R.; Kollias, A. C.; Domin, D.; Lester Jr., W. A.; Frenklach, M. *Proc. Combustion Institute* **2007,** 31, 539.
2.   Bengaard, H. S.; Norskov, J. K.; Sehested, J.; Clausen, B. S.; Nielsen, L. P.; Molenbroek, A. M.; Rostrup-Nielsen, J. R. *J. Catal.* **2002,** 209, 365.
3.   Klusek, Z.; Kozlowski, W.; Waqar, Z.; Datta, S.; Burnell-Gray, J. S.; Makarenko, I. V.; Gall, N. R.; Rutkov, E. V.; Tontegode, A. Y.; Titkov, A. N. *Appl. Surf. Sci.* **2005,** 252, 1221.
4.   Lee, S. M.; Lee, Y. H.; Hwang, Y. G.; Hahn, J. R.; Kang, H. *Phys. Rev. Lett.* **1999,** 82, 217.
5.   Liu, L.; Ryu, S. M.; Tomasik, M. R.; Stolyarova, E.; Jung, N.; Hybertsen, M. S.; Steigerwald, M. L.; Brus, L. E.; Flynn, G. W. *Nano Letters* **2008,** 8, 1965.

6.   Carlsson, J. M.; Hanke, F.; Linic, S.; Scheffler, M. *Phys. Rev. Lett.* **2009,** 102, 166104.
7.   Dreyer, D. R.; Park, S.; Bielawski, C. W.; Ruoff, R. S. *Chem. Soc. Rev.* **2010,** 39, 228.
8.   Loginova, E.; Bartelt, N. C.; Feibelman, P. J.; McCarty, K. F. *New J. Phys.* **2008,** 10, 093026.
9.   Loginova, E.; Bartelt, N. C.; Feibelman, P. J.; McCarty, K. F. *New J. Phys.* **2009,** 11, 063046.
10.  McCarty, K. F.; Feibelman, P. J.; Loginova, E.; Bartelt, N. C. *Carbon* **2009,** 47, 1806.
11.  Loginova, E.; Nie, S.; Bartelt, N. C.; Thürmer, K.; McCarty, K. F. *Phys. Rev. B* **2009,** 80, 085430.
12.  Starodub, E.; Maier, S.; Stass, I.; Bartelt, N. C.; Feibelman, P. J.; Salmeron, M.; McCarty, K. F. *Phys. Rev. B: Condens. Matter* **2009,** 80, 235422.
13.  Zangwill, A., *Physics at surfaces*. Cambridge University Press: 1988; p p. 400.
14.  Cui, Y.; Fu, Q.; Zhang, H.; Tan, D. L.; Bao, X. H. *J. Phys. Chem. C* **2009,** 113, 20365.
15.  Loginova, E.; Bartelt, N. C.; Feibelman, P. J.; McCarty, K. F. *New Journal of Physics* **2009,** 11, 063046.
16.  de la Figuera, J.; Bartelt, N. C.; McCarty, K. F. *Surf. Sci.* **2006,** 600, 4062.
17.  Madey, T. E.; Engelhardt, H. A.; Menzel, D. *Surf. Sci.* **1975,** 48, 304.
18.  Pfnur, H.; Feulner, P.; Menzel, D. *J. Chem. Phys.* **1983,** 79, 4613.
19.  Pfnur, H.; Held, G.; Lindroos, M.; Menzel, D. *Surf. Sci.* **1989,** 220, 43.
20.  Wintterlin, J.; Bocquet, M. L. *Surf. Sci.* **2009,** 603, 1841.
21.  Zhang, H.; Fu, Q.; Cui, Y.; Tan, D.; Bao, X. *J. Phys. Chem. C* **2009,** 113, 8296.
22.  Martoccia, D.; Willmott, P. R.; Brugger, T.; Bjorck, M.; Gunther, S.; Schleputz, C. M.; Cervellino, A.; Pauli, S. A.; Patterson, B. D.; Marchini, S.; Wintterlin, J.; Moritz, W.; Greber, T. *Phys. Rev. Lett.* **2008,** 101, 126102.
23.  van Gastel, R.; N'Diaye, A. T.; Wall, D.; Coraux, J.; Busse, C.; Buckanie, N. M.; zu Heringdorf, F. J. M.; von Hoegen, M. H.; Michely, T.; Poelsema, B. *Appl. Phys. Lett.* **2009,** 95.
24.  Lee, G. D.; Wang, C. Z.; Yoon, E.; Hwang, N. M.; Kim, D. Y.; Ho, K. M. *Phys. Rev. Lett.* **2005,** 95, 205501.
25.  N'Diaye, A. T.; Bleikamp, S.; Feibelman, P. J.; Michely, T. *Phys. Rev. Lett.* **2006,** 97, 215501.
26.  Pimpinelli, A.; Villain, J., *Physics of crystal growth*. Cambridge University Press: 1998.
27.  N'Diaye, A. T.; van Gastel, R.; Martinez-Galera, A. J.; Coraux, J.; Hattab, H.; Wall, D.; zu Meyer, H. F.-J.; von Horn, H. M.; Gomez-Rodriguez, J.-M.; Poelsema, B.; Busse, C.; Michely, T. *New J. Phys.* **2009**.

**Figures**

**Figure 1.** Schematic representation of graphene etching by oxygen exposure: (a) at initial stage of oxidation, carbon monomers on the metal substrate are consumed by oxygen while graphene is still at equilibrium with the monomers. The graphene etching rate is described by eq 1. (b) When the carbon monomers are depleted, oxygen might penetrate under the graphene sheet, causing its rapid etching.

**Figure 2.** Left panel: LEEM images (25 μm × 25 μm) of a graphene island on Ru(0001) at 1020 K after 0 s, 232 s, 258 s, and 274 s of exposure to $5\times10^{-9}$ Torr oxygen. Top right panel: C adatom concentration and oxygen pressure during etching. Bottom right panel: Area, perimeter, and step velocity of the graphene island.

**Figure 3.** Graphene step velocity as a function of carbon monomer concentration on Ru(0001) at 1070 K (red), 1020 K (blue), and 980 K (green). Filled circles show growth rates during C deposition [8]. Hollow circles show etching rates of graphene islands in $5\times10^{-9}$ Torr oxygen. The solid black lines are the previously published fits of the growth data to eq 1 with three parameters $B$, $c^{eq}$, and $n$.

**Figure 4.** Graphene step velocity on Ru(0001) at 1020 K during exposure to $1\times10^{-8}$, $6\times10^{-9}$, $5\times10^{-9}$, $4\times10^{-9}$ and $1\times10^{-9}$ Torr oxygen as a function of (a) C monomer concentration and (b) oxygen dose. Line is fit of growth data to eq 1 [8].

**Figure 5.** LEEM images (41 μm × 22 μm) of graphene island on Ru(0001) during exposure to $6\times10^{-8}$ Torr of oxygen at 550 K. Exposure times and oxygen doses are labeled. (Movie version is available in supplementary materials).

**Figure 6.** LEEM images (41 μm × 22 μm) of the graphene island on Ru(0001) intercalated in Figure 5 while annealing in vacuum. Annealing time and temperatures are labeled. (Movie version is available in supplementary materials).

**Figure 7.** Right panel: LEEM images (20 μm field-of-view) of two abutted graphene islands on Ir(111) during $O_2$ exposure at 1200 K. The dark island is an R0 rotational domain and the bright island is an R30 rotational domain. Top right panel: Deviation of C adatom concentration from equilibrium concentration (black) and oxygen pressure (grey) vs. time. Bottom right panel: Step velocity of the R0 (dashed black) and R30 (solid grey) graphene islands during oxygen exposure.

**Figure 8.** LEEM images (33 μm × 24 μm) during $O_2$ exposure of the graphene-covered Ir(111) at pressure $4\times10^{-9}$ Torr and temperature 1200 K. The darkest top and the brightest bottom islands are the R0 and R30 domains.

**Figure 9.** LEEM images (18.3 μm × 7.6 μm) of graphene-covered Ir(111) at 1020 K during exposure to $6\times10^{-7}$ Torr oxygen. The dark stripe on the top is graphene-free Ir(111). The bright area below is 1 ML graphene except for the dark bilayer island. Schematic cross sections show film thickness along traces marked by arrows in images.

**Figure 10.** LEEM images (20 μm field-of-view) of a graphene island on Ir(111) at 670 K during exposure to $2\times10^{-9}$ Torr oxygen. The dark island on top is R0 graphene; the bright island on bottom is R30 graphene. The striped grey area around the islands is bare, stepped substrate.

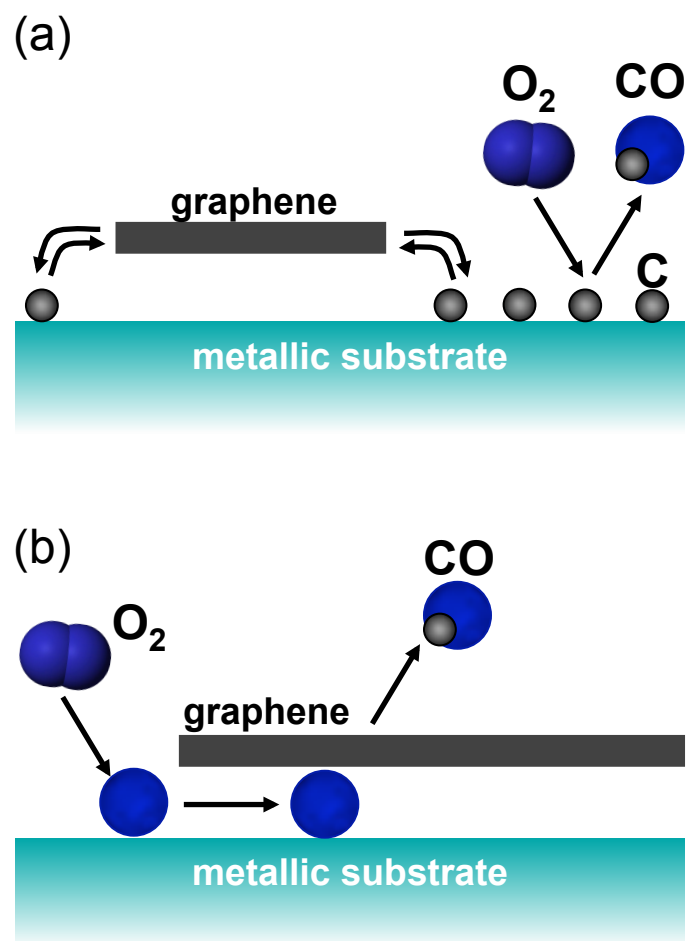

**Figure 1.**

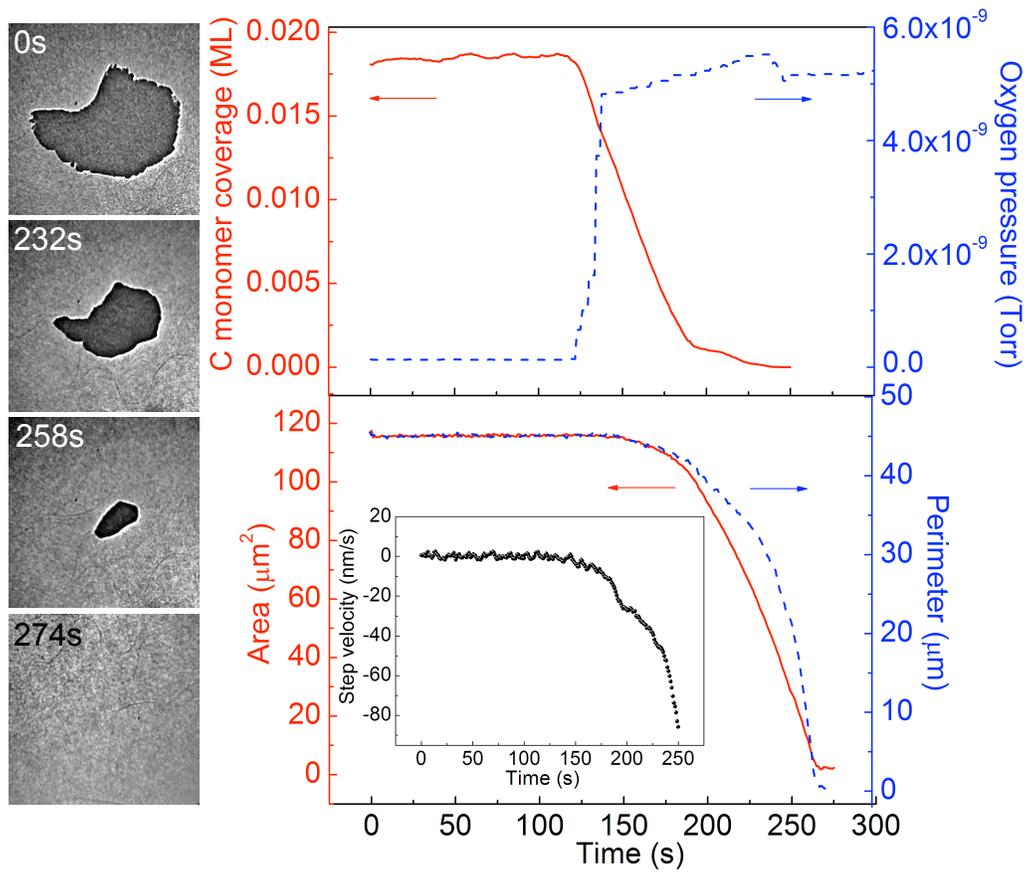

**Figure 2.**

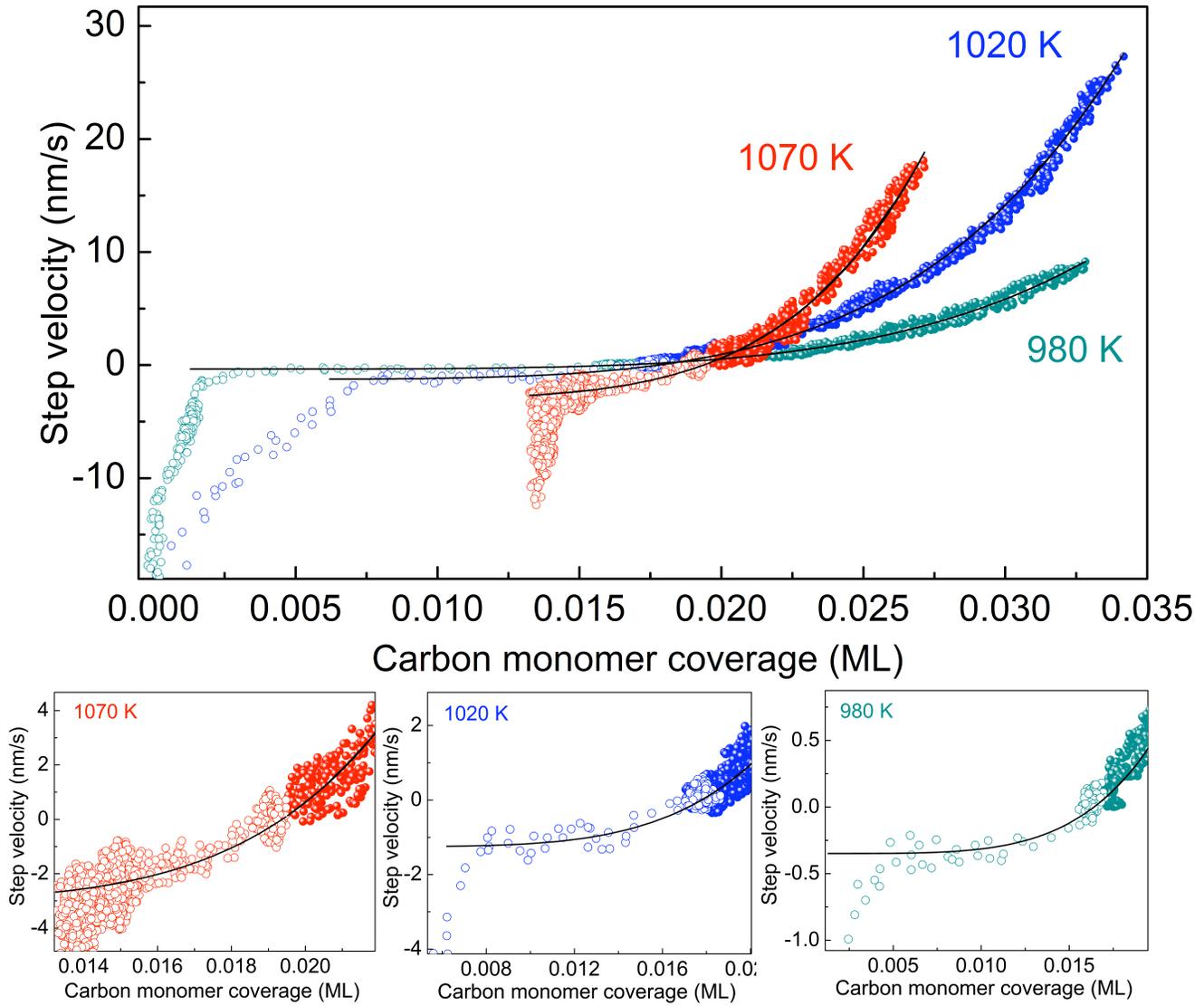

Figure 3.

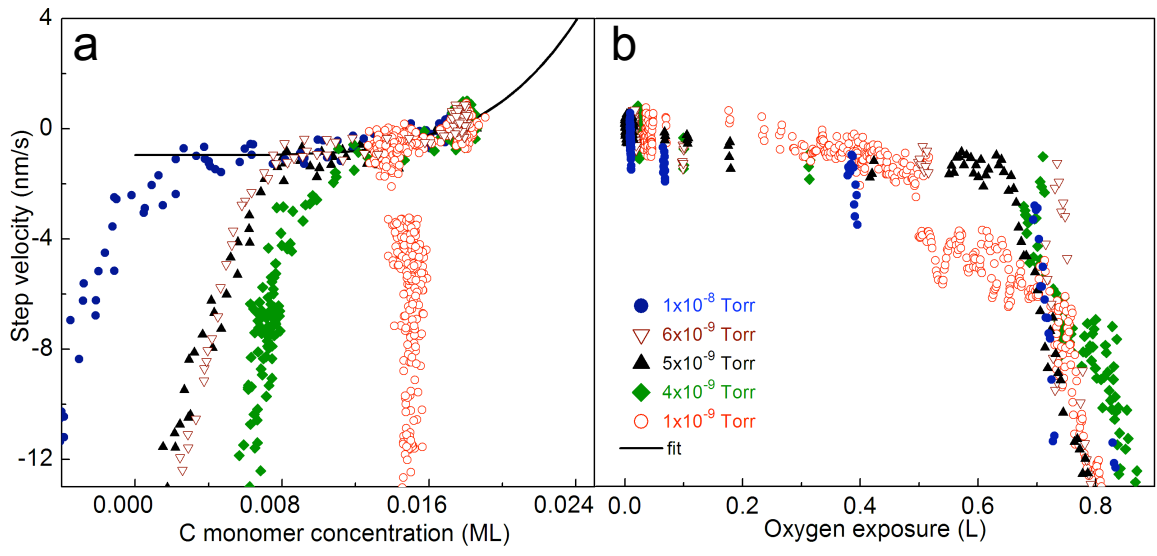

**Figure 4.**

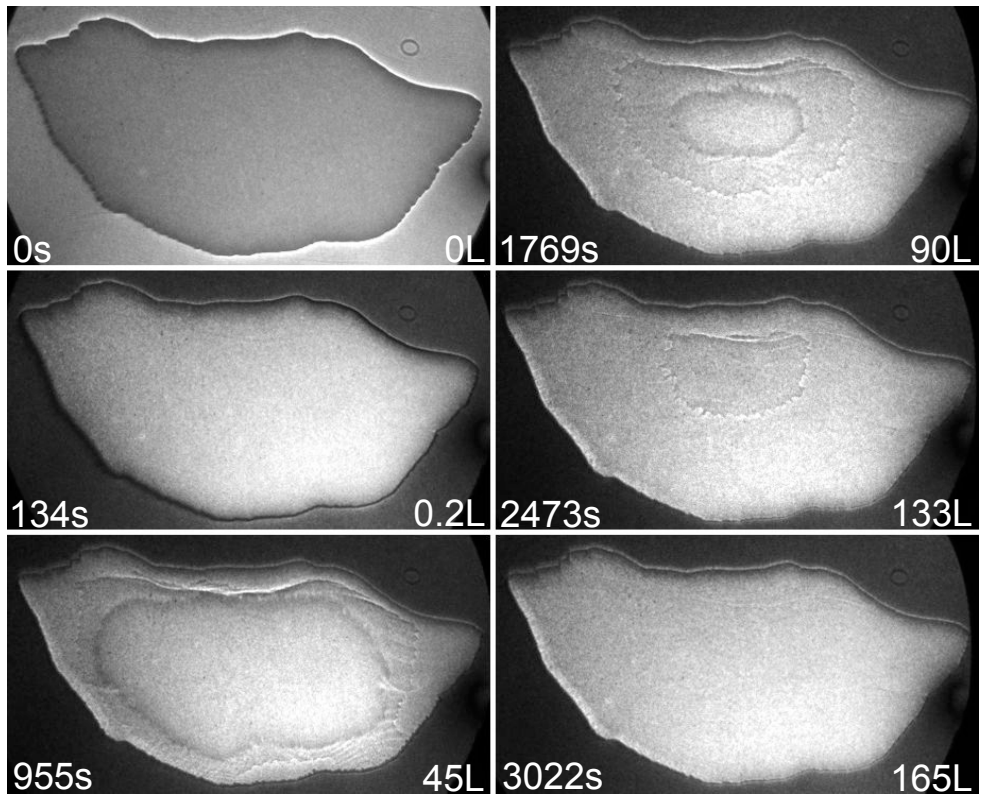

**Figure 5.**

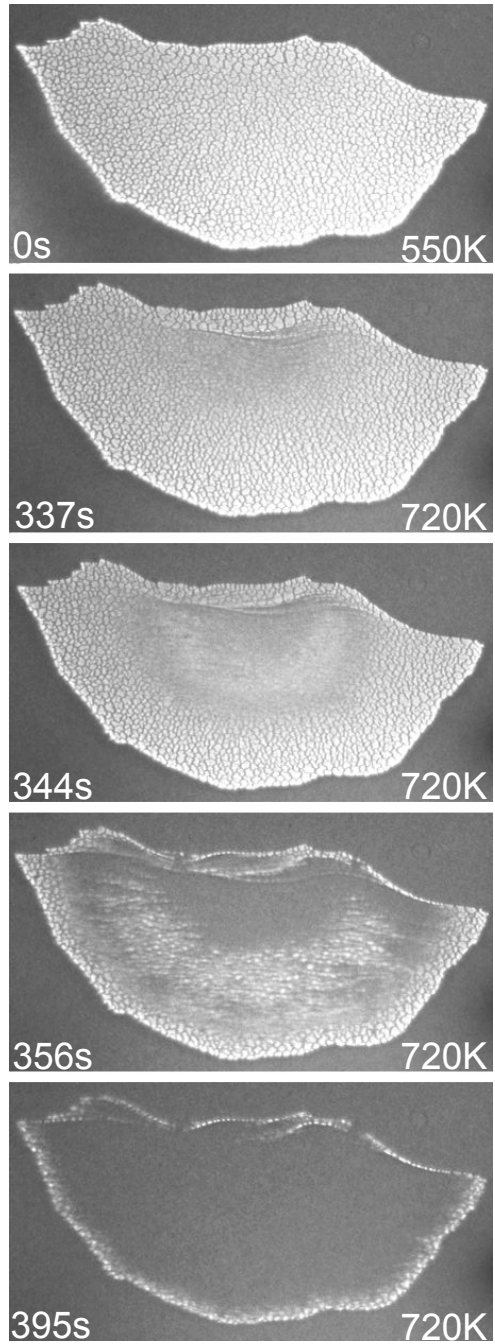

**Figure 6.**

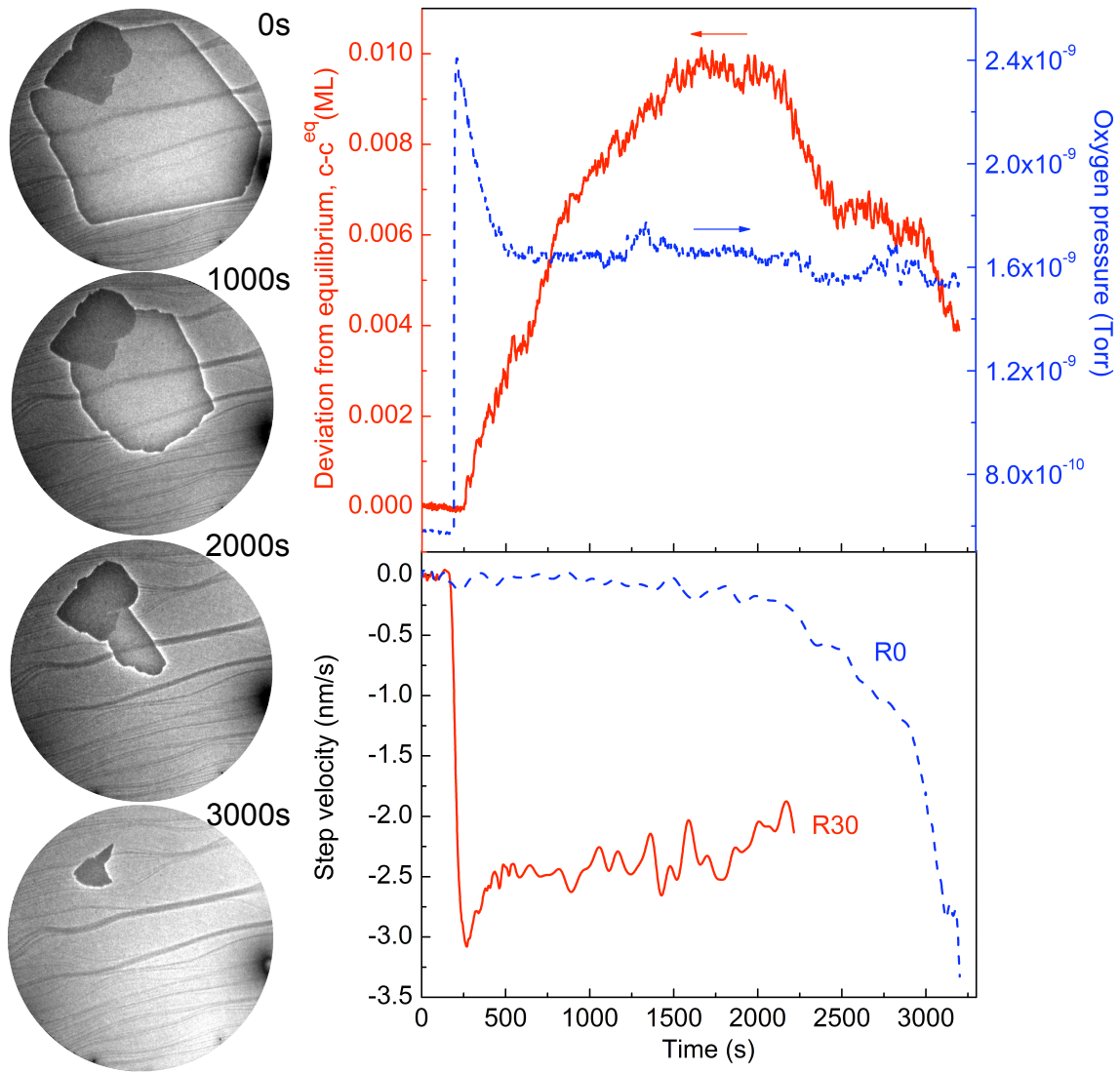

**Figure 7.**

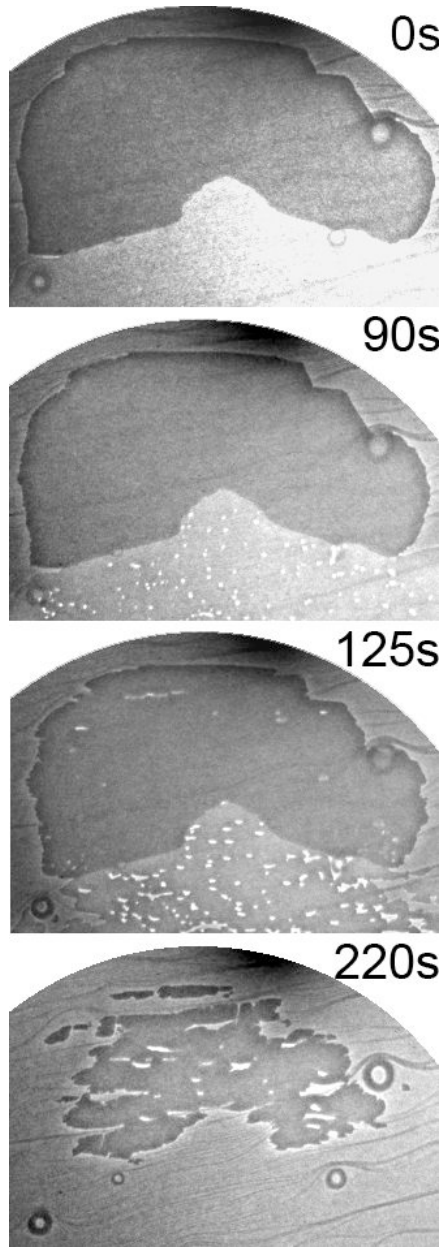

**Figure 8.**

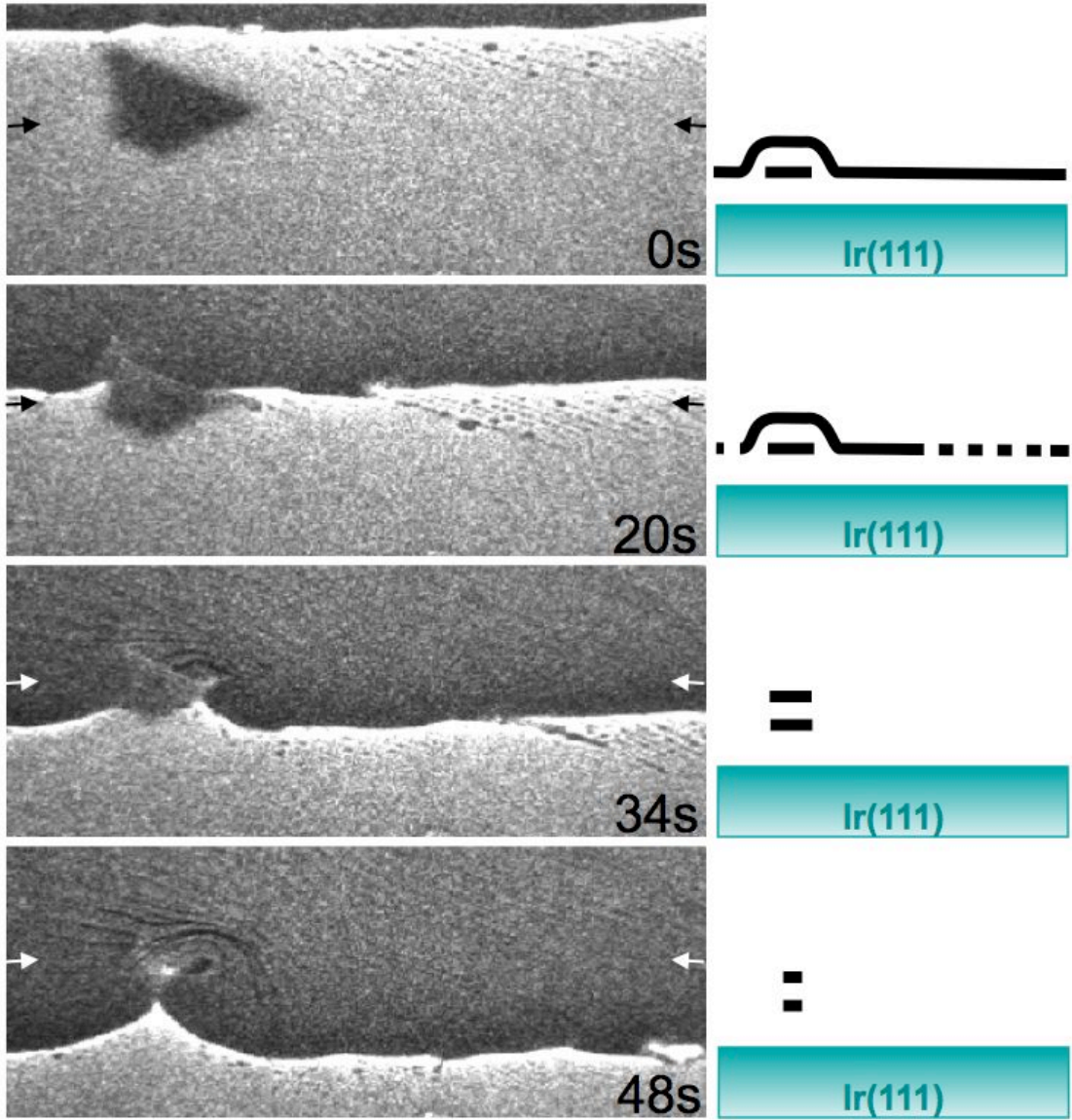

**Figure 9.**

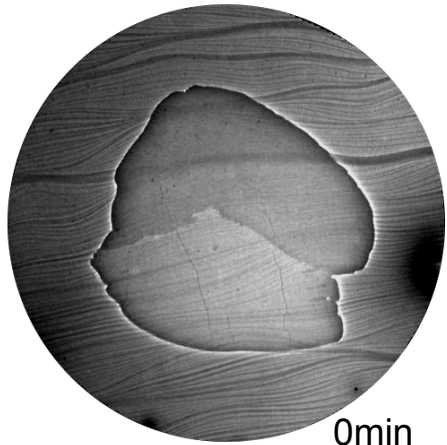
0min

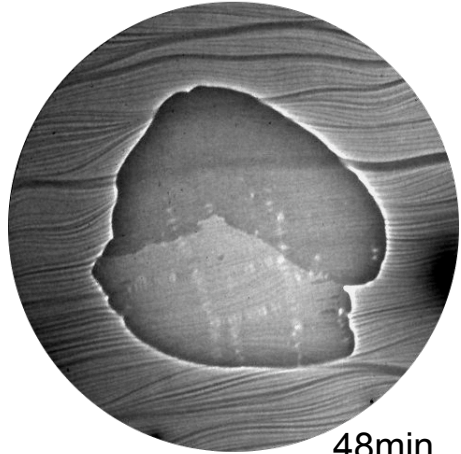
48min

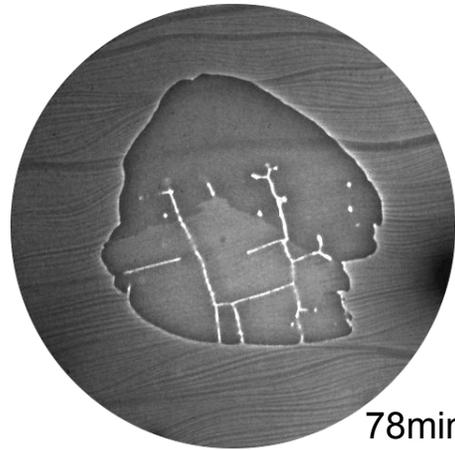
78min

**Figure 10.**

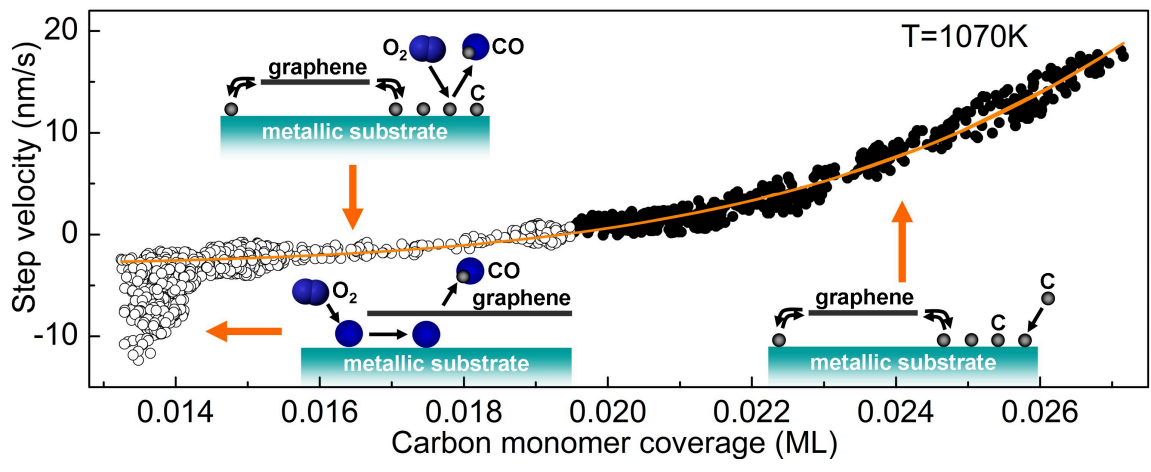

Table of contents graphic